\documentclass[iop]{emulateapj}
\usepackage{savesym}
\usepackage{amssymb}
\usepackage{amsmath}
\usepackage{hyperref}
\usepackage[]{natbib}
\usepackage{cleveref}
\usepackage{gensymb}

\newcommand{\rcs}{RCS2 J232727.6-020437}
\newcommand{\lya}{$\mathrm{Ly}\alpha$}
\newcommand{\beq}{\begin{equation}}
\newcommand{\eeq}{\end{equation}}

\def\al{\alpha}
\def\be{\beta}

\def\la{\lambda}

\def\La{\Lambda}

\def\Om{\Omega}

\def\frac#1#2{{\textstyle{{#1}\over {#2}}}}

\def\lsim{\mathrel{\rlap{\lower4pt\hbox{\hskip1pt$\sim$}}
    \raise1pt\hbox{$<$}}}
\def\gsim{\mathrel{\rlap{\lower4pt\hbox{\hskip1pt$\sim$}}
    \raise1pt\hbox{$>$}}}
\def\sqr#1#2{{\vcenter{\vbox{\hrule height.#2pt
         \hbox{\vrule width.#2pt height#1pt \kern#1pt
         \vrule width.#2pt}
         \hrule height.#2pt}}}}

\def\frac#1#2{{\textstyle{{#1}\over {#2}}}}

\def\lsim{\mathrel{\rlap{\lower4pt\hbox{\hskip1pt$\sim$}}
    \raise1pt\hbox{$<$}}}
\def\gsim{\mathrel{\rlap{\lower4pt\hbox{\hskip1pt$\sim$}}
    \raise1pt\hbox{$>$}}}
\def\sqr#1#2{{\vcenter{\vbox{\hrule height.#2pt
         \hbox{\vrule width.#2pt height#1pt \kern#1pt
         \vrule width.#2pt}
         \hrule height.#2pt}}}}

\newcommand{\bea}{\begin{eqnarray}}
\newcommand{\eea}{\end{eqnarray}}
\newcommand{\bit}{\begin{itemize}}
\newcommand{\eit}{\end{itemize}}

\def\picture #1 by #2 (#3){
  \vbox to #2{
    \hrule width #1 height 0pt depth 0pt
    \vfill
    \special{picture #3} 
    }
  }

\def\scaledpicture #1 by #2 (#3 scaled #4){{
  \dimen0=#1 \dimen1=#2
  \divide\dimen0 by 1000 \multiply\dimen0 by #4
  \divide\dimen1 by 1000 \multiply\dimen1 by #4
  \picture \dimen0 by \dimen1 (#3 scaled #4)}
  }

\begin{document}

\title{\rcs: An Efficient Cosmic Telescope at $z=0.6986$}

\author {A.~Hoag\altaffilmark{1}, M.~Brada\v{c}\altaffilmark{1}, K.-H.~Huang\altaffilmark{1}, R.E.~Ryan Jr\altaffilmark{2}, K.~Sharon\altaffilmark{3}, T.~Schrabback\altaffilmark{4}, K.B.~Schmidt\altaffilmark{5}, B.~Cain\altaffilmark{1}, A.H.~Gonzalez\altaffilmark{6}, H.~Hildebrandt\altaffilmark{4}, J.L.~Hinz\altaffilmark{7}, B.C.~Lemaux\altaffilmark{8}, A.~von der Linden\altaffilmark{9,10,11}, L.M.~Lubin\altaffilmark{1}, T.~Treu\altaffilmark{12}, D.~Zaritsky\altaffilmark{13}}

\altaffiltext{1}{Department of Physics, University of California, Davis, CA, 95616, USA}
\altaffiltext{2}{Space Telescope Science Institute 3700 San Martin Dr. Baltimore MD, 21218, USA}
\altaffiltext{3}{Department of Astronomy, University of Michigan, 1085 S. University Ave, Ann Arbor, MI 48109, USA}
\altaffiltext{4}{Argelander-Institut f\"ur Astronomie, Auf dem H\"ugel 71, D-53121 Bonn, Germany}
\altaffiltext{5}{Department of Physics, University of California, Santa Barbara, CA, 93106-9530, USA}
\altaffiltext{6}{Department of Astronomy, University of Florida, Gainesville, FL, 32611, USA}
\altaffiltext{7}{MMT Observatory, 933 N Cherry Ave, Tucson AZ, 85721, USA}
\altaffiltext{8}{Aix Marseille Universit\'{e}, CNRS, LAM (Laboratoire d'Astrophysique de Marseille) UMR 7326, 13388 Marseille, France}
\altaffiltext{9}{Department of Physics, Stanford University, 382 Via Pueblo Mall,  Stanford, CA  94305-4060, USA}
\altaffiltext{10}{Dark Cosmology Centre,  Niels Bohr Institute, University of Copenhagen, Juliane Maries Vej 30, 2100 Copenhagen {\O}, Denmark}
\altaffiltext{11}{Kavli Institute for Particle Astrophysics and Cosmology, Stanford University, 452 Lomita Mall, Stanford, CA  94305-4085, USA}
\altaffiltext{12}{Department of Physics and Astronomy, UCLA, Los Angeles, CA, 90095-1547, USA}
\altaffiltext{13}{Steward Observatory, University of Arizona, 933 N. Cherry Ave., Tucson, AZ, 85721, USA}

\begin{abstract}

We present a detailed gravitational lens model of the galaxy cluster {\rcs}. Due to cosmological dimming of cluster members and ICL, its high redshift ($z=0.6986$) makes it ideal for studying background galaxies. Using new ACS and WFC3/IR HST data, we identify 16 multiple images. From MOSFIRE follow up, we identify a strong emission line in the spectrum of one multiple image, likely confirming the redshift of that system to $z=2.083$. With a highly magnified ($\mu\gtrsim2$) source plane area of $\sim0.7$ arcmin$^2$ at $z=7$, {\rcs} has a lensing efficiency comparable to the Hubble Frontier Fields clusters. We discover four highly magnified $z\sim7$ candidate Lyman-break galaxies behind the cluster, one of which may be multiply-imaged. Correcting for magnification, we find that all four candidates are fainter than $0.5 L_{\star}$. One candidate is detected at ${>10\sigma}$ in both Spitzer/IRAC [3.6] and [4.5] channels. A spectroscopic follow-up with MOSFIRE does not result in the detection of the Lyman-alpha emission line from any of the four candidates. From the MOSFIRE spectra, we place median upper limits on the Lyman-alpha flux of $5-14 \times 10^{-19}\, \mathrm{erg \,\, s^{-1} cm^{-2}}$ ($5\sigma$). 

\end{abstract}

\section{Introduction}\label{sec:intro}

It has been decades since the first evidence that clusters of galaxies could be used as cosmic telescopes. \citet{soucail90} originally proposed the idea to study magnified background galaxies. Several large programs have recently been developed to capitalize on the magnification gain. Some of these programs include the Cluster Lensing And Supernova survey with Hubble (CLASH, PI Postman: HST-GO-12065, \citealp{postman12}), the Grism Lens-Amplified Survey from Space (GLASS, PI Treu: HST-GO-13459, \citealp{schmidt14}), and the Hubble Frontier Fields\footnote{http://www.stsci.edu/hst/campaigns/frontier-fields} (HFF, PI Lotz: Program ID 13495).  

These programs were established largely to better understand one of the remaining frontiers in cosmology, the epoch of reionization, referring to the transition of neutral to ionized hydrogen in the intergalactic medium (IGM) at redshifts above 6. It is likely that star-forming galaxies at these redshifts reionized the IGM (\citealp{robertson10}, \citealp{bouwens12}), though the duration of the process is still debated (\citealp{robertson15}). The continued search with the Wide Field Camera 3 (WFC3) on board the \textit{Hubble Space Telescope} (HST) and with future instruments on upcoming telescopes such as the \textit{James Webb Space Telescope} and 30-m class telescopes is likely to reveal the exact role of the first galaxies during this epoch. 

\begin{figure*}  
        \centering
        \includegraphics[width=\linewidth]{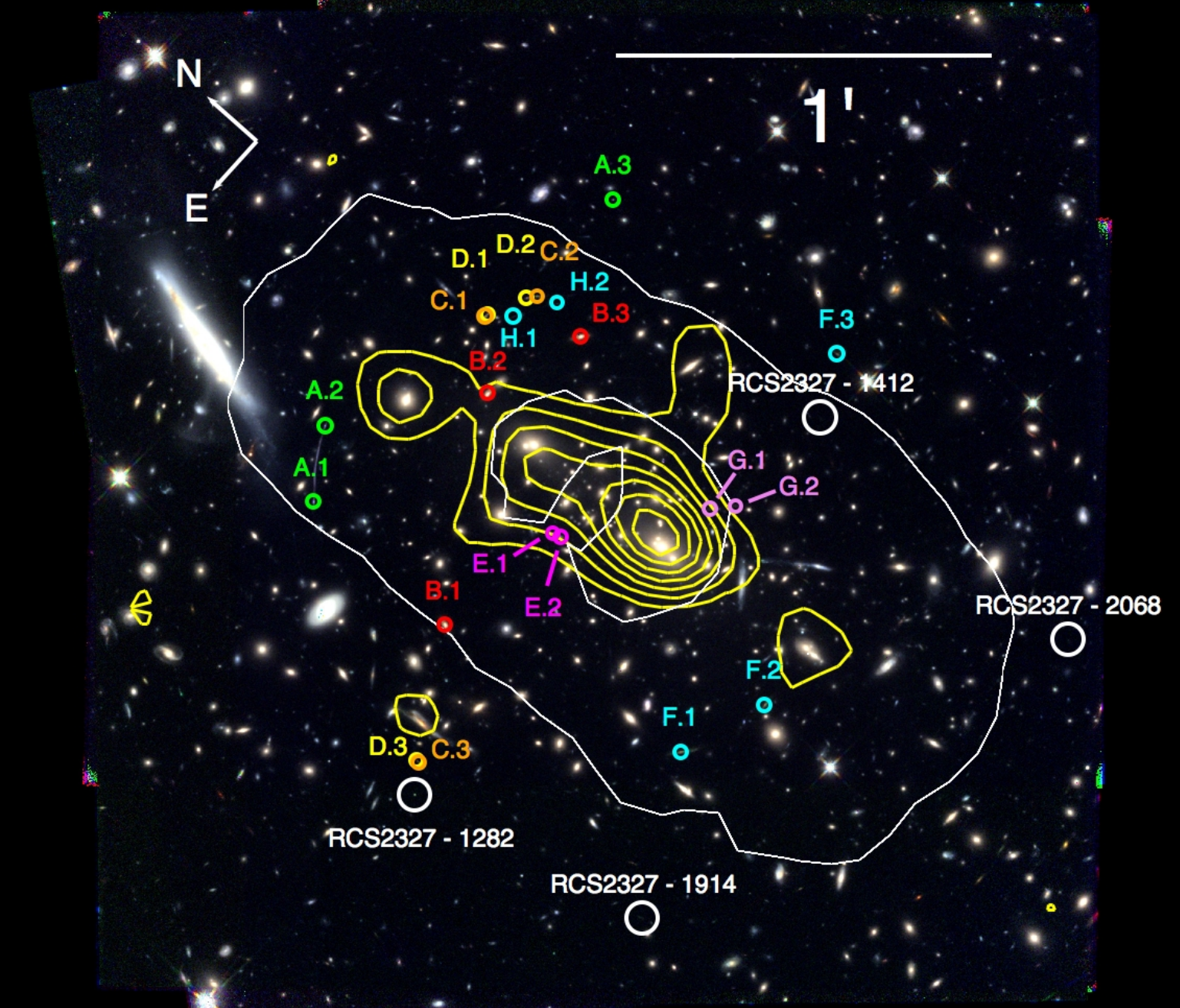}                
	\caption{Multiple image systems used to constrain the lens model (color-coded by system). The large white circles enclose the four primary F814W-dropout candidates. The critical curve (white line) at $z=7$ from our best-fit lens model is shown. Also shown are stellar mass contours (yellow) for cluster members obtained from Spitzer/IRAC data. The color image is a combination of HST filters: F098M, F125W, and F160W.  }
	\label{fig:rgbarcs}
\end{figure*}

Here we present a glimpse behind one of the best cosmic telescopes, the galaxy cluster {\rcs} (hereafter RCS2327). RCS2327 is the richest and most massive cluster discovered by the RCS-2 survey (\citealp{rcs2}, \citealp{sharon15}). Original imaging data from this survey revealed a giant arc (Fig.~\ref{fig:rgbarcs}). Einstein radii (\citealp{gralla11}), Sunyaev-Zel'dovich masses (\citealp{gralla11}, \citealp{menanteau13}) and a strong lensing mass (\citealp{sharon15}) have since been estimated, all corroborating the extreme mass of the cluster ($\mathrm{M_{200}} \sim 3\times10^{15} \mathrm{M_{\cdot}}$). \citet{sharon15} measured the redshift of the cluster to be $z=0.6986$ with optical spectroscopy of several hundred galaxies. The cluster is at higher redshift than all of the HFF clusters and 24 of the 25 CLASH clusters. High redshift cosmic telescopes have a distinct advantage over their lower redshift counterparts of the same mass distribution: cosmological dimming of cluster members and intra-cluster light (ICL) decreases the obscured fraction of the telescope FOV.  


Key parameters that influence lensing efficiency include the mass, geometrical configuration, ellipticity of the total mass distribution, amount of substructure in the cluster (\citealp{meneghetti03}) and along the line of sight (\citealp{wong12}). An efficient cosmic telescope exhibits large magnification factors ($\mu>2$) over a large area of the cluster. The magnification comes with a price: in order to recover intrinsic properties of magnified objects, knowledge of the local magnification is required. 

In this paper, we present a gravitational lens model of RCS2327 and the first search for $z\sim7$ candidates behind it. Section~\ref{sec:data} contains a summary of our new imaging and spectroscopic data. In Section~\ref{sec:lensmodel}, we show all multiple images identified so far in RCS2327, including the 16 new images presented in this work. We present four sub-$L_{\star}$ $z\sim7$ Lyman-break galaxy (LBG) candidates discovered in our new HST data (Section~\ref{sec:dropselect}). Finally, we put upper limits on the flux and the {\lya} rest frame equivalent width (EW) of the four dropout candidates from our spectroscopic follow-up with Keck/MOSFIRE (Section~\ref{sec:limits}). 

Throughout this paper, we assume a flat $\La$CDM cosmology with $h=0.7$, $\Om_m = 0.3$, $\Om_{\La} = 0.7$, and $H_0 = 100 \textrm{h km s$^{-1}$ Mpc$^{-1}$}$. All magnitudes refer to the AB system (\citealp{oke74}).

\section{Observations and Data Reduction}{\label{sec:data}}

\subsection{Imaging data}{\label{sec:imagedata}}

The original optical imaging data of RCS2327 were taken with HST+ACS Wide Field Channel (WFC) in F435W (PI: Gladders, HST-GO-10846). Those data are presented in \citet{sharon15}. We obtained new optical and near-IR imaging data with HST+ACS and WFC3/IR as part of the Spitzer Ultra Faint Survey Program (SURFSUP). SURFSUP is a joint Spitzer (PI: Bradac, PID: 90009)  and HST (PI: Bradac, HST-GO-13177) Exploration Science Spitzer program. The observation and data reduction procedures for our HST and Spitzer data are detailed in \citet{bradac14} and \citet{ryan14}. 


\begin{deluxetable}{cccc}
\tablecolumns{4}
\tablecaption{\bf Photometric data summary} 
\tablehead{\colhead{Filter} & \colhead{$\mathrm{t_{exp}}$ (s) } & \colhead{$\mathrm{mag_{lim}} (1\sigma$)} & \colhead{HST cycle} }
\medskip 
\startdata
F435W & 4208 & 28.7 & 15 \\ 
F814W & 5626 & 28.1 & 20 \\ 
F098M & 6835 & 28.0 & 20 \\ 
F125W & 6635 & 27.4 & 20 \\ 
F160W & 9247 & 27.6 & 20 \\ 
\vspace{-0.2cm}\\ \hline
\vspace{-0.1cm} \\
IRAC [3.6] & $10^{5}$ & -- & -- \\
IRAC [4.5] & $10^{5}$ & -- & -- 
\enddata
\tablecomments{\label{table:hstfilts} Photometric Filters used to observe RCS2327, their exposure times and $1\sigma$ limiting magnitudes. The limiting magnitudes are obtained from the upper limits of the flux error distribution found from artificial source simulations. IRAC exposure times are approximated from portions of the coverage maps overlapping with the WFC3/IR FOV.  } 
\end{deluxetable}


Exposure times and limiting magnitudes for the HST filters employed in this paper are shown in Table~\ref{table:hstfilts}. To measure HST limiting magnitudes, we mimic the photometric procedure described in Section~\ref{sec:dropselect} for our F814W-dropout candidates. We first create 5000 simulated point sources with $m=25$, placing them in random locations in a combined $\mathrm{F098M} + \mathrm{F125W} + \mathrm{F160W} $ detection image. We then run SExtractor (\citealp{sextractor}) in dual-image mode on each of the HST filters in Table~\ref{table:hstfilts}, finding that it detects $\sim$4600 of the 5000 simulated sources. In the five individual HST filters, where there are no sources added, we measure the isophotal flux at the locations of each simulated source in the detection image. The standard deviation of this flux distribution, converted to a magnitude, is the $1\sigma$ limiting magnitude (mag\_lim) quoted in Table~\ref{table:hstfilts}. We assess the positional dependence of the limiting magnitude in each filter. In the vicinity of each dropout candidate, the limiting magnitudes differ very little from the global limiting magnitudes reported. We therefore use the global limiting magnitudes in Section~\ref{sec:dropselect}. 

As part of the SURFSUP program, we also obtained infrared imaging data from the Spitzer/IRAC [3.6] and [4.5] bands (\citealp{fazio04}). Due to the larger point-spread-function in these bands relative to HST+ACS and WFC3/IR, a different approach is necessary to conduct photometry. As a quick summary of our IRAC photometry procedure (\citealp{bradac14}, Huang et al. 2015, in preparation), we use the software package TFIT (\citealp{laidler07}) and the segmentation map in WFC3/IR F160W as the high-resolution prior for mixed-resolution photometry. TFIT uses the flux within the isophotal aperture of each object as the high-resolution template, convolves it with a transformation kernel that matches the PSFs in F160W and IRAC bands, and for each object solves the flux that optimizes the match to the IRAC pixel values. We model all detected objects in F160W within a 20$''$ by 20$''$ box centered at each high-$z$ candidate to account for blending in IRAC. In the cases where object confusion is too severe to obtain good fits (either due to imperfect PSF models or color gradients between F160W and IRAC), we derive the 3$\sigma$ flux limits for each high-$z$ candidate from simulations (Huang et al. 2015, in preparation). The simulations are similar to those done to measure the HST limiting magnitudes. 

\subsection{Ground-Based Spectroscopic Data}
 
Preexisting spectroscopic data obtained by \citet{sharon15} confirm the redshifts of multiple image systems A\&B. New spectroscopic observations of RCS2327 took place on 2013 Dec 15, Dec 17 and Dec 18 (UTC) in Y and H-band using the \textit{Multi-Object Spectrometer for Infrared Exploration}  (MOSFIRE, \citealp{mosfire}, \citealp{mclean12}) on the Keck I telescope. Observations totaled 3 hours in each band spread evenly over the three nights, with mostly sub-arcsecond seeing (Table~\ref{table:mosobs}).

We created a multi-object slit-mask using the MOSFIRE Automatic GUI-based Mask Application\footnote{\url{http://www2.keck.hawaii.edu/inst/mosfire/magma.html}}. On Dec 15 and Dec 17 we used the same mask to observe RCS2327. On Dec 18, we used a secondary mask to target different objects in filler slits. However, we observed all dropout candidates on all three nights using the same slit configuration. We employed a $2.5''$ ABBA nod pattern consisting of 180 (120) second individual exposures in our Y-band (H-band) observations of RCS2327. These data were reduced and combined using the publicly available MOSFIRE data reduction pipeline (DRP\footnote{\url{https://code.google.com/p/mosfire/}}). The reduction pipeline differences, stacks, and rectifies the nodded images, creating 2-dimensional signal and inverse variance spectra for each individual slit. Spectra from individual nights were reduced separately due to varying observing conditions.

To extract 1-dimensional spectra from the 2-dimensional DRP products, we define a constant (in the spatial direction) aperture size. The size is chosen to enclose $99.5$\% of the light smeared by the seeing\footnote{We assume a circular Gaussian profile for the seeing PSF in this calculation.}. Pixels within the aperture are summed at each point along the spectral axis to produce a 1-dimensional spectrum. Flux calibration is done by comparing a 1-dimensional extracted spectrum of a standard star we observed to a model spectrum of the same spectral type. Model spectra are obtained from the CALSPEC\footnote{http://www.stsci.edu/hst/observatory/crds/calspec.html} database. Before comparison, we scale the model spectrum to the apparent magnitude of the observed standard, then correct it for the airmass and galactic extinction of our observations. The ratio of the scaled, corrected model spectrum to the observed telluric spectrum represents the sensitivity function of the telescope.

To obtain calibrated science spectra of each dropout candidate, we first extract a 1-dimensional spectrum from the 2-dimensional signal spectrum produced by the DRP. We then correct this spectrum for airmass and extinction. The corrected spectrum is multiplied by the sensitivity function obtained as described above. We combine spectra from the three separate nights of observations to improve the signal-to-noise ratio ($S/N$) of the spectra. Following an approach described by \citet{finkelstein13}, we scale the noise spectra so that the standard deviation of the $S/N$ spectra equals 1. This ensures that the noise spectra accurately reflect the errors in the signal spectra. $S/N$ spectra are smoothed to the FWHM of an unresolved emission line, which is $\sim3$ pixels throughout Y and H-band (\citealp{mclean12}). 

The seeing during our observations is measured in J-band alignment images of stars taken prior to each observation in Y-band. We use the IRAF routine imexamine to measure the FWHM of five alignment stars in each image\footnote{IRAF is distributed by the National Optical Astronomy Observatories, which are operated by the Association of Universities for Research in Astronomy, Inc., under cooperative agreement with the National Science Foundation.}. The seeing value for each Y-band observation reported in Table~\ref{table:mosobs} is the mean of the FWHM of the light distribution measured from five different alignment stars in the same frame taken before each set of science exposures. H-band science exposures were always taken directly after Y-band science exposures. Consequently, we calculate the seeing in H-band by comparing the FWHM of the continuum trace of a bright object in both bands. We account for slit losses due to the seeing after calibrating our science spectra.


\begin{deluxetable}{lccc}
\tablecolumns{4}
\tablecaption{\bf MOSFIRE spectroscopic data summary \label{table:mosobs} } 
\tablehead{\colhead{Night} & \colhead{band} & \colhead{$\mathrm{t_{exp}}$ (s)} & \colhead{ Seeing ($"$) } }
\medskip
\startdata
2013 Dec 15 & Y & 3579 &  0.75 \\ 
                     & H & 3340 & 0.65 \\
\\
2013 Dec 17 & Y & 3579 & 0.88 \\ 
  		     & H & 3340 & 0.92 \\ 
\\
2013 Dec 18 & Y & 3579 & 1.20 \\ 
  		     & H & 3340 & 0.86 \\ 
\\[-0.5cm]
\enddata
\tablecomments{Seeing was measured in observations of alignment stars in J-band directly preceding science exposures.} 

\end{deluxetable}


\section{Lens Model} {\label{sec:lensmodel}}

The lens modeling method we employ, SWUnited (\citealp{bradac05}, \citealp{bradac09}), constrains the gravitational potential within a galaxy cluster field via $\chi^2$ minimization. It takes as input a basic initial model for the potential. A $\chi^2$ is then calculated from strong and weak gravitational lensing data on an adaptive, pixelated grid over the potential ($\psi_k$) established by the initial model. Areas closer to the core(s) of the mass distribution and in the vicinity of multiple images are reconstructed with higher resolution. By iteratively solving a set of linearized equations satisfying $\partial \chi^2 / \partial \psi_k = 0$, a minimum $\chi^2$ is found. Derivative lensing quantity maps, such as convergence ($\kappa$), shear ($\gamma$) and magnification ($\mu$), are produced from the best-fit potential map. 

\subsection{Multiple Image Systems}{\label{sec:arcs}}


\begin{deluxetable}{ccccc}[H]
\tablecolumns{5}
\tablecaption{\bf Multiple Images \label{table:arcs} } 
\tablehead{\colhead{ID} & \colhead{$\alpha_{J2000}$} & \colhead{$\delta_{J2000}$} & \colhead{$z$ } & \colhead{F160W } }
\startdata
\vspace{0.2cm}
A.1 & 351.87405 & -2.0643629 & 2.98\tablenotemark{a} & 23.00 \\ 
\vspace{0.2cm}
A.2 & 351.87116 & -2.0624877 & 2.98\tablenotemark{a} & 23.47 \\ 
\vspace{0.2cm}
A.3 & 351.85518 & -2.0652714 & $3.00^{+0.18}_{-0.18}$ & 24.07 \\ 
\vspace{0.2cm}
B.1 &  351.87420 & -2.0723517 & 1.42\tablenotemark{a} & 20.60 \\ 
\vspace{0.2cm}
B.2 & 351.86529 & -2.0669059 & 1.42\tablenotemark{a} & 20.63 \\  
\vspace{0.2cm}
B.3 & 351.86065 & -2.0682697	& 1.42\tablenotemark{a} & 20.77 \\  
\vspace{0.2cm}
C.1 & 351.86283 & -2.0645228 & 2.083\tablenotemark{b} & 22.75 \\ 
\vspace{0.2cm}
C.2 & 351.86062 & -2.0656345 & -- & 22.96 \\ 
\vspace{0.2cm}
C.3 & 351.87950 & -2.0755689 & $2.35^{+0.31}_{-0.82}$ & 24.97 \\ 
\vspace{0.2cm}
D.1 & 351.86267 & -2.0645370 & $2.13^{+0.07}_{-0.09}$ & 22.75 \\ 
\vspace{0.2cm}
\hspace{0.15cm}D.2\tablenotemark{c} & 351.86099 & -2.0653389 & -- & 23.63 \\ 
\vspace{0.2cm}
D.3 & 351.87950 & -2.0754385 & $2.10^{+0.11}_{-0.13}$ & 24.16 \\ 
\vspace{0.2cm}
\hspace{0.3cm}E.1\tablenotemark{d}\tablenotemark{e} & 351.86797 & -2.0731903 & $2.85^{+0.15}_{-0.12}$ & 22.61 \\ 
\vspace{0.2cm}
\hspace{0.35cm}E.2\tablenotemark{d}\tablenotemark{e}\tablenotemark{f} & 351.86785 & -2.0735543 & -- & -- \\ 
\vspace{0.2cm}
F.1 & 351.87138 & -2.0838888 & $1.69^{+0.07}_{-0.07}$ & 22.22 \\ 
\vspace{0.2cm}
F.2 & 351.86734 & -2.0852578 & $1.63^{+0.07}_{-0.08}$ & 22.53 \\ 
\vspace{0.2cm}
F.3 & 351.85361 & -2.0772366 & $1.56^{+0.11}_{-0.11}$ & 23.37 \\ 
\vspace{0.2cm}
\hspace{0.25cm}G.1\tablenotemark{c}\tablenotemark{e} & 351.86261 & -2.0775600 & -- & 24.66 \\ 
\vspace{0.2cm}
\hspace{0.25cm}G.2\tablenotemark{c}\tablenotemark{e} & 351.86165 & -2.0784200 & -- & 23.77 \\ 
\vspace{0.2cm}
H.1 & 351.86199 & -2.0654498 &  $2.17^{+0.54}_{-1.58}$ & 25.57 \\ 
\hspace{0.1cm}H.2 & 351.86025 & -2.0665151 & $2.35^{+0.28}_{-0.29}$ & 25.03 
\enddata
\tablenotetext{a}{Spectroscopic redshift confirmed by \citet{sharon15}.}
\tablenotetext{b}{Likely spectroscopic redshift from this work.} 
\tablenotetext{c}{Contamination from cluster members/ICL did not permit a reliable photo-z measurement.}
\tablenotetext{d}{May be part of system A. }
\tablenotetext{e}{Not used in the lens model. }
\tablenotetext{f}{Detected as same object as E.1 in SExtractor segmentation map. }

\tablecomments{Column $z$ list the photometric redshift with 68\% confidence intervals or the spectroscopic redshift, if available. } 
\end{deluxetable}

 
The strong lensing signal is estimated using positions of multiple image systems, which can be identified in images of the cluster. To determine if objects are multiple images of the same source, we require that the measured photometric redshift (photo-$z$) of each image falls within the $68\%$ confidence limits of all other images in the system.  We compute photo-$z$ distributions for multiple images using Le Phare (\citealp{lephare1}, \citealp{lephare2}). We use the Le Phare keyword Z\_ML, the median of the maximum likelihood (ML) distribution, to estimate the redshift of the system. Another requirement is that the morphologies of the images are consistent, accounting for the parities, distortion and magnification predicted by the best-fit lens model. 

Two multiple image systems (A \& B) were previously identified in imaging data obtained during the RCS-2 survey and were spectroscopically confirmed to be at redshifts $z=2.98$ and $z=1.42$, respectively (\citealp{sharon15}). These two systems provide the strongest constraints for our model. We identify 15 multiple images belonging to 6 new systems that obey the criteria outlined above. We also identify a candidate counter-image (A.3) to the giant arc. We measure its photo-$z$ (Table~\ref{table:arcs}), and we find that the spec-$z$ measured by \citet{sharon15} is within the $68\%$ confidence limits. A.3 also possesses the expected morphology of a counter-image to the giant arc.


\begin{figure}
\centering
	\includegraphics[width=8cm,height=6cm ]{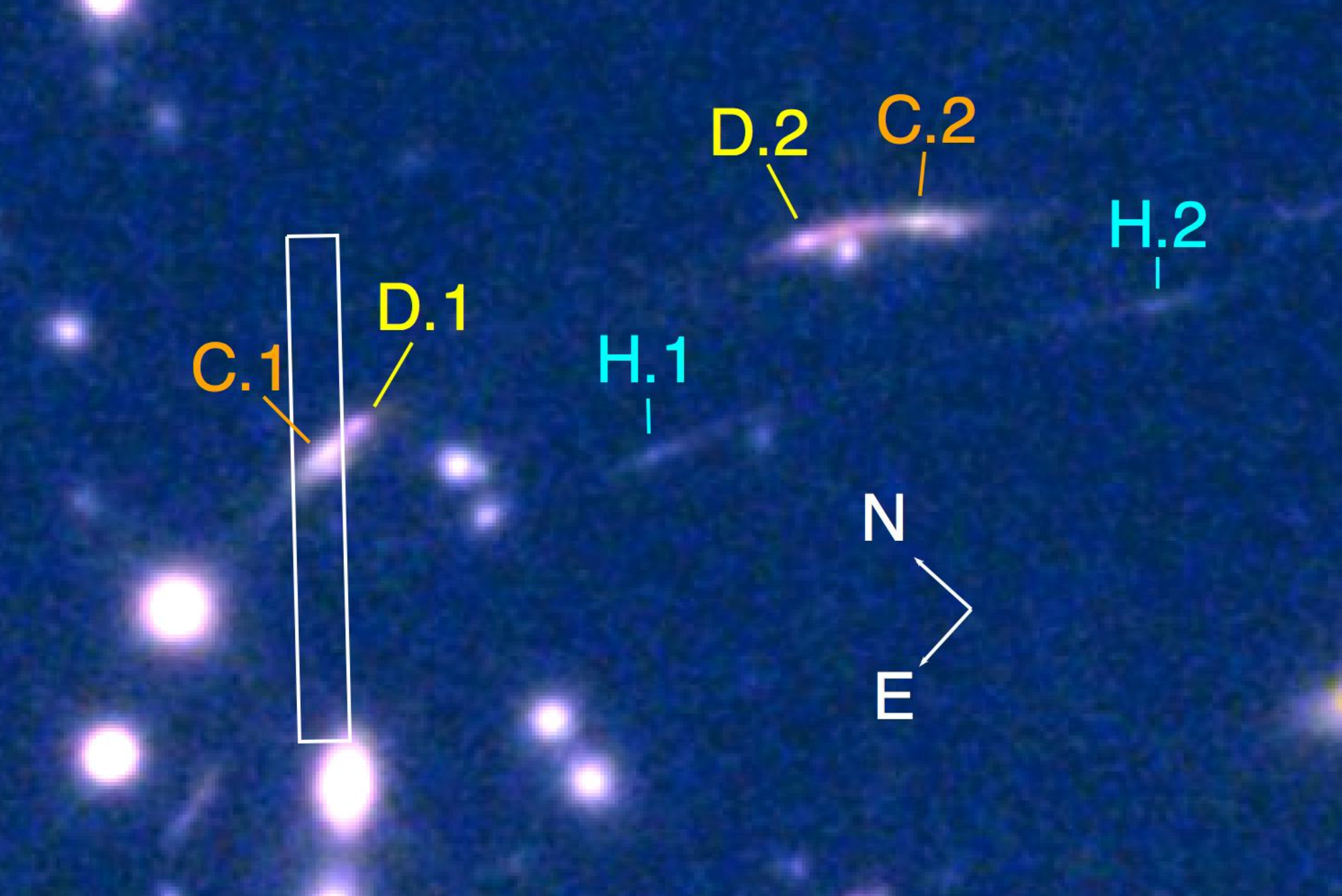}
	\caption{$12''\times19''$ close-up of multiple image systems C and D, showing the configuration of the MOSFIRE slit ($0.7 ''\times 7 ''$) used to target image C.1 (white rectangle). The color image is a combination of HST filters: F098M, F125W, and F160W.}
	\label{fig:system_C_rgb}
\end{figure}



\begin{figure*}%
\centering
\includegraphics[width=\linewidth]{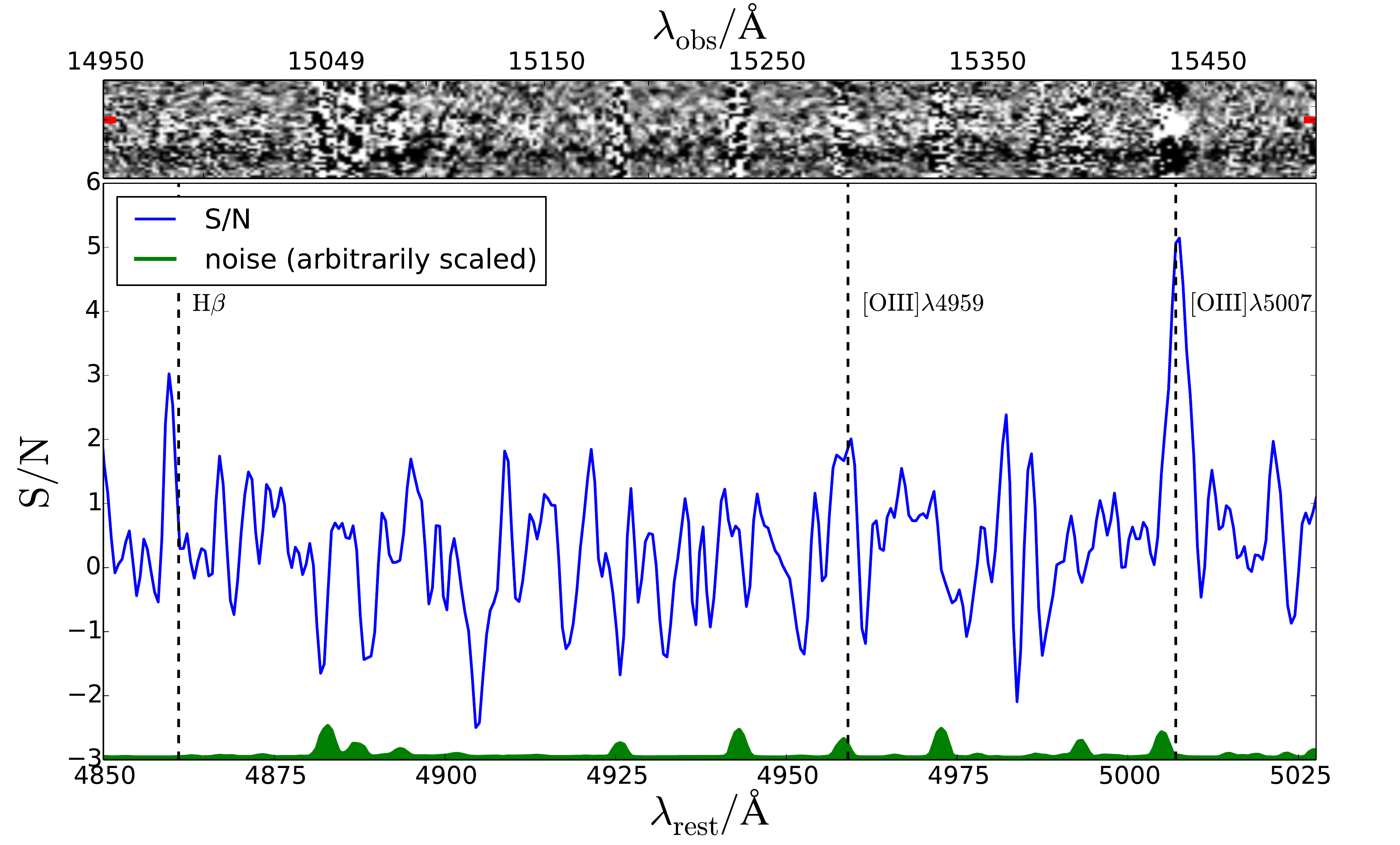}
\caption{Top panel: a portion of the 2-dimensional H-band reduced spectrum of multiple image C.1. A strong emission line is seen at $\lambda_{\mathrm{obs}} = 15436$ {\AA}, which is likely [OIII]$\la$5007 at redshift $z=2.0828 \pm 0.0004$. The red horizontal ticks at the left and right edges of the spectrum indicate the expected vertical position in the slit from broadband HST data. A skyline can be seen just blue-ward of the emission line, limiting the precision of the redshift measurement. The portion of the spectrum shown is $6.7''$ in the vertical direction. Bottom panel: 1-dimensional signal-to-noise ratio ($S/N$) spectrum extracted from the 2-dimensional spectrum. The spectrum is shown at rest frame wavelengths assuming redshift $z=2.083$. Vertical dotted lines indicate the wavelengths of predicted emission lines in this scenario. The line flux ratios are consistent with the line flux ratios for star-forming galaxies (\citealp{nagao06}). Shown below the spectrum is an arbitrarily-scaled pure noise spectrum (green). The peaks of the noise spectrum are due to strong sky emission lines contaminating our spectrum. The $S/N$ spectrum was smoothed using a gaussian kernel of $\sigma = 1$ MOSFIRE pixel. }
\label{fig:Hbandlines}
\end{figure*}

\begin{figure}
        \centering
        \includegraphics[trim= 2cm 0cm 0cm 0cm, width=\linewidth,height=7cm]{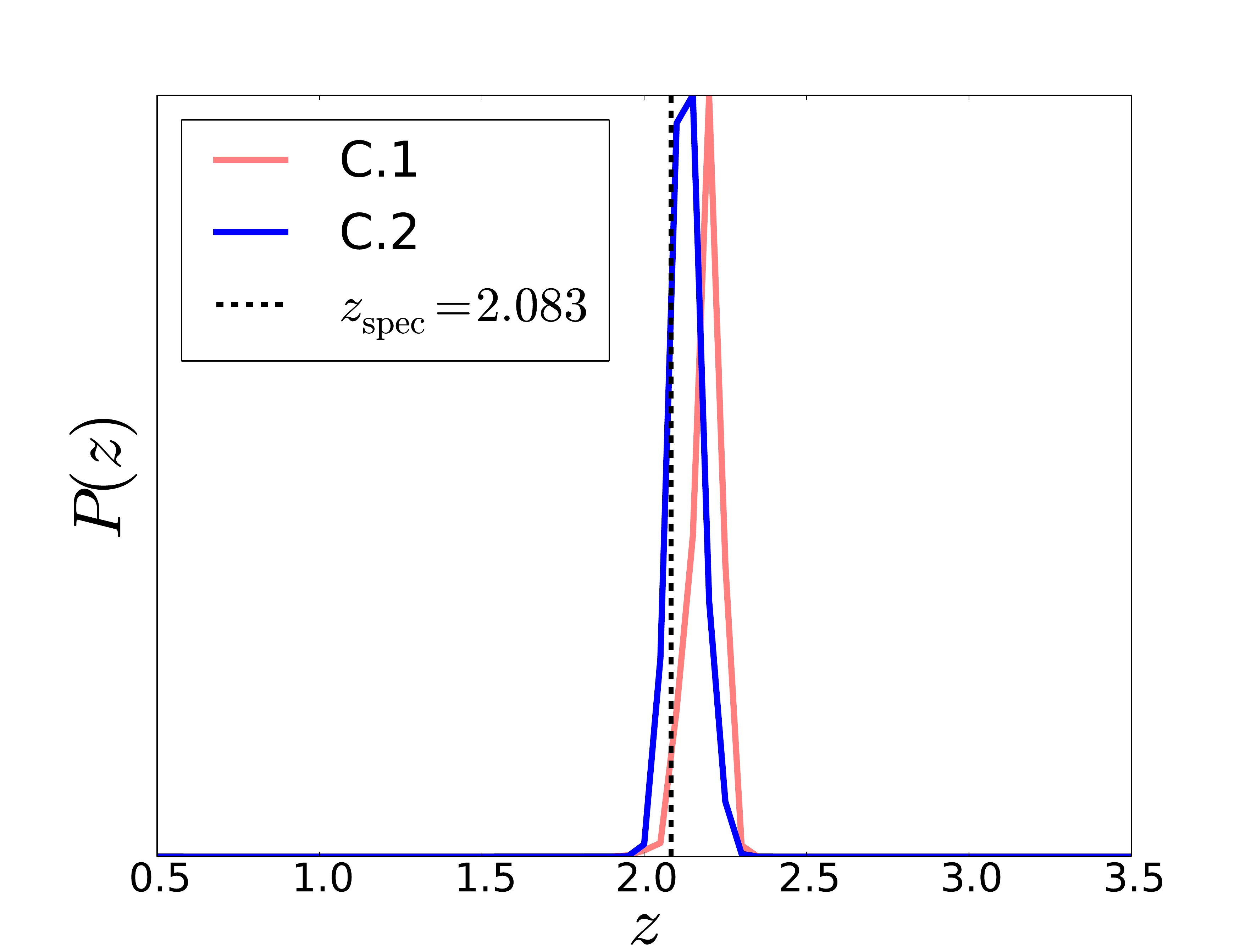}
        \caption{Photometric redshift distributions for images C.1 and C.2. Together, the distributions put a tight constraint on the redshift of the system, in excellent agreement with the spectroscopic redshift (dotted black line) suggested by the emission line detected in the MOSFIRE H-band spectrum of C.1. Note that there are no other probability peaks outside of the range shown.}
        \label{fig:system_C_pzs}
\end{figure}

During our MOSFIRE observations of RCS2327, we targeted image C.1 (Fig.~\ref{fig:system_C_rgb}). We identify a single H-band emission line at $15436$ {\AA} (Fig.~\ref{fig:Hbandlines}). The photo-$z$'s of C.1 and C.2 are precise and consistent with each other (Fig.~\ref{fig:system_C_pzs}). The two best solutions to the single emission line wavelength based on photo-$z$ are [OIII]$\la$5007 at $z=2.083$ or H$\be$ at $z=2.176$, where the uncertainties are dominated by the presence of the nearby sky line. The line cannot be [OIII]$\la$4959 at $z=2.112$ because we would detect [OIII]$\la$5007, the brighter of the [OIII]$\la\la$4959,5007 doublet, at 15581 {\AA} with high significance; the line flux ratio $\textrm{F([OIII]}\la5007) / F(\textrm{[OIII]}\la4959)$ is always 3 in the absence of instrumental or atmospheric effects. The ratio, $R = \textrm{F([OIII]}\la5007) / F(\textrm{H}\be)$, for star forming galaxies is highly dependent on the gas metallicity. We compute R for a range of metallicities, $0.1 Z_{\odot} < Z <  Z_{\odot}$, using a polynomial fit determined by \citet{nagao06}. We find that for metallicities $Z\lesssim0.5 Z_{\odot}$, $ R \gtrsim 3 $. For larger metallicities, $0.5 Z_{\odot} < Z < Z_{\odot}$, the line strengths become comparable. If this were the case, then we would expect to see both lines in the spectrum. However, such large metallicities are improbable at $z\sim2$. Therefore, it is much more likely that the line is [OIII]$\la$5007 at $z=2.083$.

It is difficult the measure the line flux of the observed emission line at 15436 {\AA} due to the bright sky line with which it is blended. We extract a 1-dimensional $S/N$ spectrum of C.1 in H-band and show the positions of expected emission lines in the case of the bright emission line being [OIII]$\la$5007. The peak $S/N$ of the bright emission line at 15436 is 5.1. A sky line heavily contaminates the spectrum at the location of $\textrm{[OIII]}\la4959$, at which point the peak $S/N$ is 2. However, we do not expect a significant detection in the absence of contamination for the reason described above. At the expected wavelength of H$\be$, we see a feature with peak $S/N = 3$, though offset by $\sim1 \AA$ in the rest frame. This feature is in a relatively clean part of the spectrum, but does not have the convincing negative residuals at the expected locations due to our dithering pattern (Section~\ref{sec:data}). We expect H$\be$ to be less or equally significant to $\textrm{[OIII]}\la4959$ for metallicities in the range $0.1 Z_{\odot} < Z < 0.5 Z_{\odot}$. We conclude that the two features at the expected locations of H$\be$ and $\textrm{[OIII]}\la4959$ are consistent with the noise, as expected from the predicted line ratios.

All other emission line scenarios are strongly prohibited by the photometric redshift. However, we examine a few other scenarios. If the emission line at 15436 {\AA} were H$\al$ at $z=1.352$, no other strong lines would be expected in either band. Similarly, if the line were part of the [OII]$\la\la$3726-3729 doublet at $z=3.142$, no other strong lines would be visible in either band. The observed line would have to be the 3729 line, in which case the 3726 line would be fully obscured by the sky line. Because we lack a second line to confirm the spectroscopic redshift, we cannot completely rule out either the H$\al$ or the [OII]$\la\la$3726-3729 scenario.

Because multiple image systems C and D are lensed to approximately the same positions in the image plane, they must be at approximately the same redshift. The photometric redshift of D.1 (Table~\ref{table:arcs}) is consistent with the spectroscopic redshift of C.1, as expected. D.2 is blended with a cluster member, so we were unable to reliably measure its photometric redshift. Systems C and D appear to be merging, creating a visible color gradient (Fig.~\ref{fig:system_C_rgb}). Using the best-fit lens model, we predict one counter-image for each of the two systems. We identify two blended counter-images (C.3 and D.3) that display a similar color gradient. The photometric redshifts of C.3 and D.3 are consistent with the spectroscopic redshift obtained for C.1.

Photometric redshift uncertainty is a source of statistic error in our lens model. When available, independent measurements of the redshift of multiple images belonging to the same systems can reduce the statistical uncertainty, and, under certain conditions, mitigate catastrophic redshift errors. We do not include the photometric redshift error in our error budget. Instead, we treat the most probable redshift given by the joint photometric redshift distribution of each system as a spectroscopic redshift in our modeling. 

A potentially large source of systematic error in strong lensing analyses is the misidentification of multiple images. While we require multiple conditions to be satisfied in order for images to be considered part of the same system, we cannot rule out the possibility of misidentification for systems lacking spectroscopic confirmation. Deep spectra needed for redshift confirmation are expensive and often lacking for the majority of systems in strong lensing analyses (e.g. \citealp{bradac08}, \citealp{zitrin09}, \citealp{merten11}). 


\subsection{Weak Lensing Galaxies}

We generate a weak lensing galaxy shape catalog from the ACS F814W data using the KSB formalism (\citealp{kaiser95}). We largely follow \citet{schrabback10}, but additionally apply the pixel-based algorithm for the correction of charge-transfer inefficiency from \citet{massey14}, a revised weak lensing weighting scheme based on the measured rms ellipticity as a function of magnitude. We remove galaxies with \mbox{$S/N<10$}, where \mbox{$S/N$} is defined as the AUTO flux divided by the AUTO flux error measured in the SExtractor shape catalog. See Schrabback et al. (2015, in preparation) for further details. 

Red cluster members are removed from the shape catalog by identifying the red sequence. The identification is done in the $\mathrm{F814W} -\mathrm{F098M}$ vs. F098M plane.  We isolate the red sequence via 
\beq 20 < Y_{098} < 27 \nonumber \eeq
\beq 1.54 - 0.05 \cdot Y_{098} < I_{814} - Y_{098} < 1.75 - 0.05 \cdot Y_{098}, \nonumber \eeq
where $Y_{098}$ and $I_{814}$ are ISO magnitudes in the F098M and F814W HST filters, respectively. ISO magnitudes were measured then scaled to the F160W AUTO magnitude (see Section~\ref{sec:dropselect}). 

After removal of the red sequence members, 715 galaxies remain within the lensing field we consider (Section~\ref{sec:magmetric}). Each galaxy is weighted by $1 / \textrm{rms}_\textrm{e}^2$, where $\textrm{rms}_\textrm{e}$ is the rms of the measured ellipticity. Galaxies are assigned a photometric redshift equal to the median photometric redshift, $z=1.18$, of the selected weak lensing galaxy sample. \citet{applegate14} show that using photo-$z$ point estimates for weak lensing galaxies leads to large biases in the inferred cluster masses. However, the signal normalization in our lens modeling code is dominated by strong lensing. For the same reason, we are not concerned with removing the residual blue cluster members that normally dilute the weak lensing signal.

\subsection{Magnification Maps} \label{sec:magmetric}
We construct magnification maps of RCS2327 in a $4'\times4'$ field centered at the location of the brightest cluster galaxy (BCG; $\alpha_{J2000}, \delta_{J2000} = 23^{\mathrm{h}}27^{\mathrm{m}}8^{\mathrm{s}}.646, -02	\degree04'37.86$). The lensing field size is chosen to be larger than the FOV of ACS WFC, the widest of of our HST observations, to avoid edge effects in the mass reconstruction. We provide the lens modeling code with an initial model of non-singular isothermal ellipsoids (NIEs) centered at the light peaks of the BCGs. The input NIE parameters are velocity dispersion, core radius, ellipticity and orientation. We first attempt to model the cluster using only two NIEs in our initial model. However, we are unable to reproduce all of the strong lensing data without the addition of more cluster member halos. We find the best fit to the data starting from an initial model of five NIEs. We determine the best-fit run by the strong lensing source plane ${\chi_{\mathrm{SL}}^2}$. In the best-fit run, we find $\chi_{\mathrm{SL}}^2 / \mathrm{N_{images}} = 2.2$, where $\mathrm{N_{images}}$ is the total number of multiple images input to the model. Magnification maps are obtained from the best-fit reconstruction of the potential, which is smoothed using a circular Gaussian kernel before extracting magnification information.  


\begin{figure*}
        \centering
\includegraphics[width=0.8\linewidth]{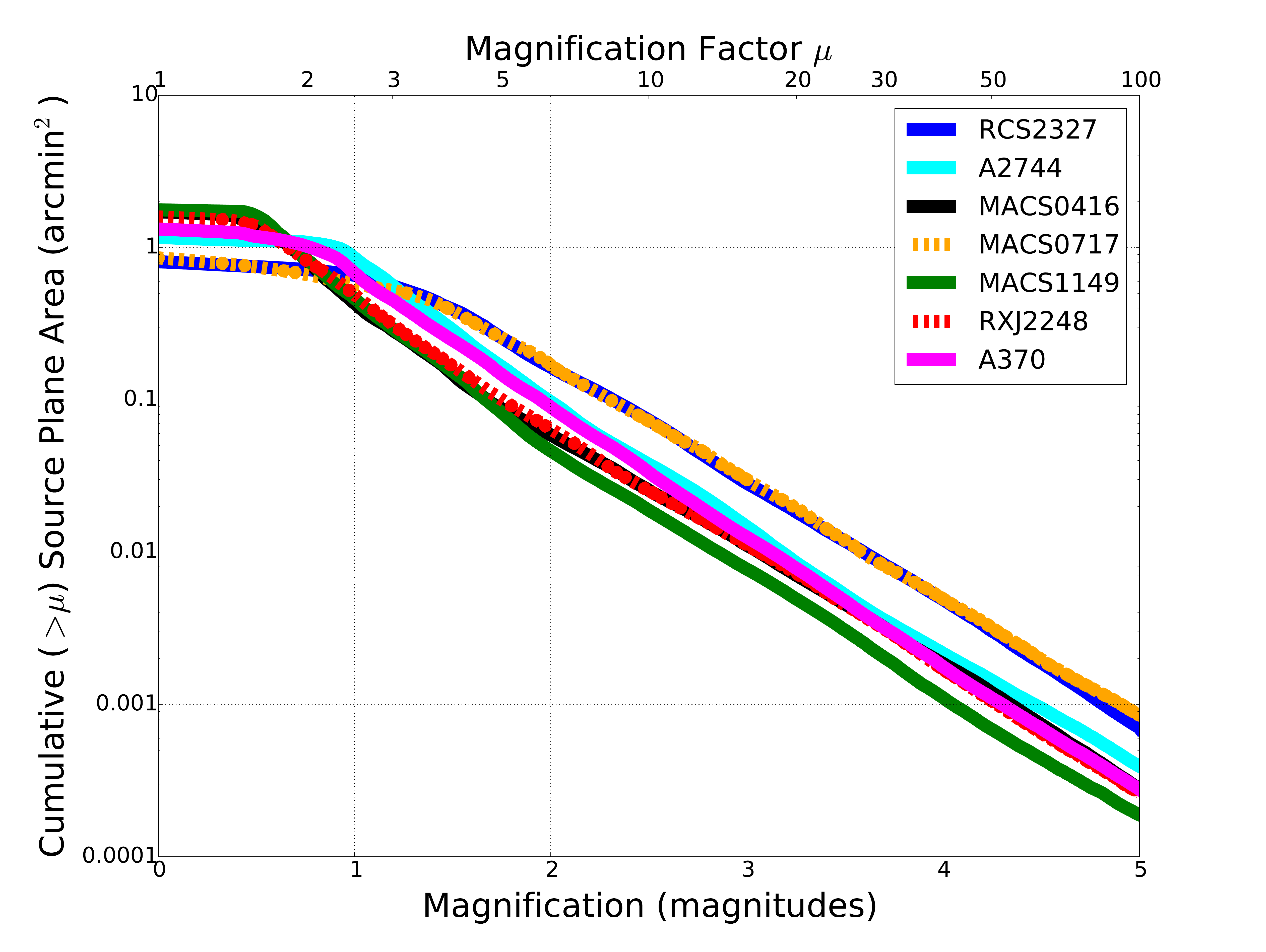}
		\caption{Cumulative source plane area versus magnification at $z=7$. RCS2327 is compared to our submitted models of the Frontier Fields. The image plane area selected for the comparison is the same for all clusters. For each cluster, we select a $\sim4.5$ arcmin$^2$ field centered at the center of the F160W pointing. For large values of magnification ($\mu \gtrsim4$), RCS2327 magnifies a larger area than $5/6$ of the HFF clusters. RCS2327 and MACS0717, one of the most massive galaxy clusters known \citep{limousin12}, follow similar curves. } 
		\label{fig:cumarea}
\end{figure*}


We compare the magnification properties of RCS2327 to a sample of the most powerful known lenses, the HFF clusters. Fig.~\ref{fig:cumarea} shows the cumulative source plane area at $z=7$ as a function of the magnification for RCS2327 and the six HFF clusters. The image plane area ($\sim4.5$ arcmin$^2$) that is lensed back to the source plane for this calculation is approximately the size of the WFC3/IR FOV and is the same for each cluster. RCS2327 is most similar in magnification properties to MACS0717, yet with rounder critical curves which makes it more suitable for observations with WFC3/IR. For values of magnification $\mu \gtrsim 4$, RCS2327 magnifies a larger area than $5/6$ of the HFF clusters. The Frontier Fields magnification maps we use to make the comparison are created from our version 1 models submitted as part of a special STScI call to map the Frontier Fields using ancillary data\footnote{\url{http://www.stsci.edu/hst/campaigns/frontier-fields/Lensing-Models}}. 

Magnification error maps are obtained by bootstrap resampling the weak lensing catalog. For each resampling, the same initial model and run parameters are used. This is done 100 times to improve statistics. Confidence limits were obtained by taking the central 68 of the 100 sample points. We note that the error maps that are generated only account for statistical uncertainties. 

\subsection{Stellar mass map}
The Spitzer/IRAC $3.6\,\mu$m image samples close to rest-frame $K$-band for the cluster, so we use the $3.6\,\mu$m fluxes from cluster members to approximate the cluster stellar mass distribution. We first select the red sequence cluster members brighter than the 25th mag in F814W from the color-magnitude and color-color diagrams following the procedure in \cite{richard14}. We select a total of 311 bright cluster members for their stellar mass distribution.

To create an image with $3.6\,\mu$m flux from cluster members only, we first create a mask with value 1 for pixels that belong to cluster members in the F160W image and 0 otherwise. We then convolve the mask with the $3.6\,\mu$m PSF to match the IRAC angular resolution, set the pixels below 10\% of the peak value to zero, and resample the mask onto the IRAC pixel grid. We obtain the $3.6\,\mu$m map of cluster members by setting all IRAC pixels not belonging to cluster members to zero and smooth the final stellar mass map with a two-pixel wide Gaussian kernel.

\section{Dropout Selection}{\label{sec:dropselect}}
\begin{figure*}
        \centering
\includegraphics[trim= 0mm 10mm 0mm 10mm,width=0.8\linewidth]{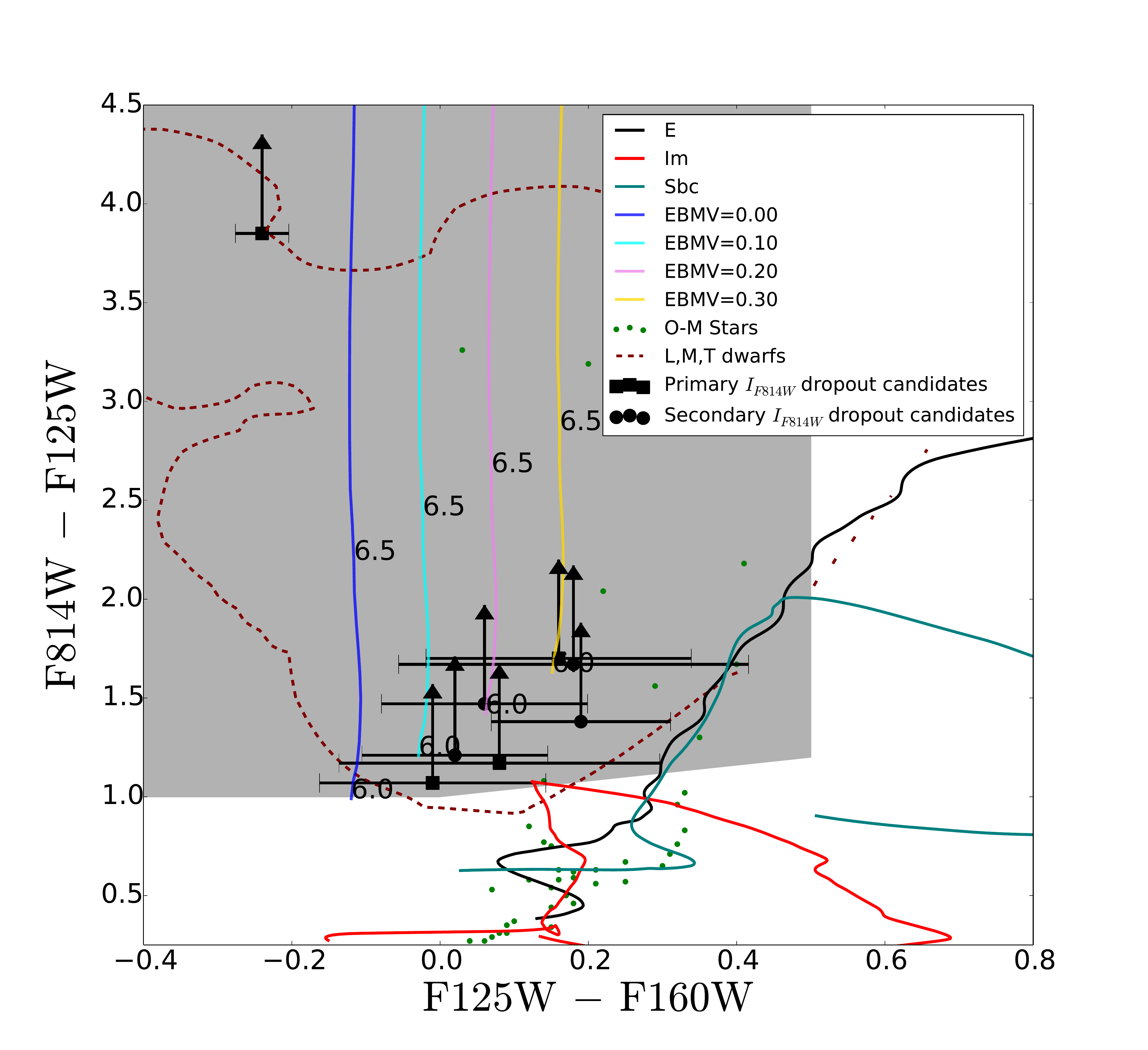}
	\caption{Four primary (black squares) and four secondary (black circles) F814W-dropout candidates. The grey shaded region is the dropout selection window defined in~(\ref{selection1}). Because all dropout candidates were not detected in F814W, vertical errors are shown as limits (black arrows). The color tracks for $z>6$ starburst galaxies including varying amounts of dust are shown in blue, cyan, magenta and yellow. The fact that all but one dropout candidate falls near the $z=6.0$ points on the starburst color tracks is an indication of the limited dynamic range in $\mathrm{F814W} - \mathrm{F125W}$ color. The outlying dropout candidate is RCS2327-1282. Also shown are possible contaminants. Color tracks for elliptical (E, black), irregular (Im, red), and Sbc (teal) galaxies were obtained from the galaxy libraries in BPZ (\citealp{bpz}). Colors for O-M stars (green points) were obtained from the Pickles library (\citealp{pickles98}). Also shown is a single density contour enclosing the portion of the diagram occupied by L,T, and M dwarfs obtained from the A. Burgasser SpeX compilation (\url{http://pono.ucsd.edu/~adam/browndwarfs/spexprism/}). }
	\label{fig:dropcolors}
\end{figure*}



\begin{figure*}%
\centering
\includegraphics[trim=10mm 10mm 10mm 15mm, clip,width=\linewidth]{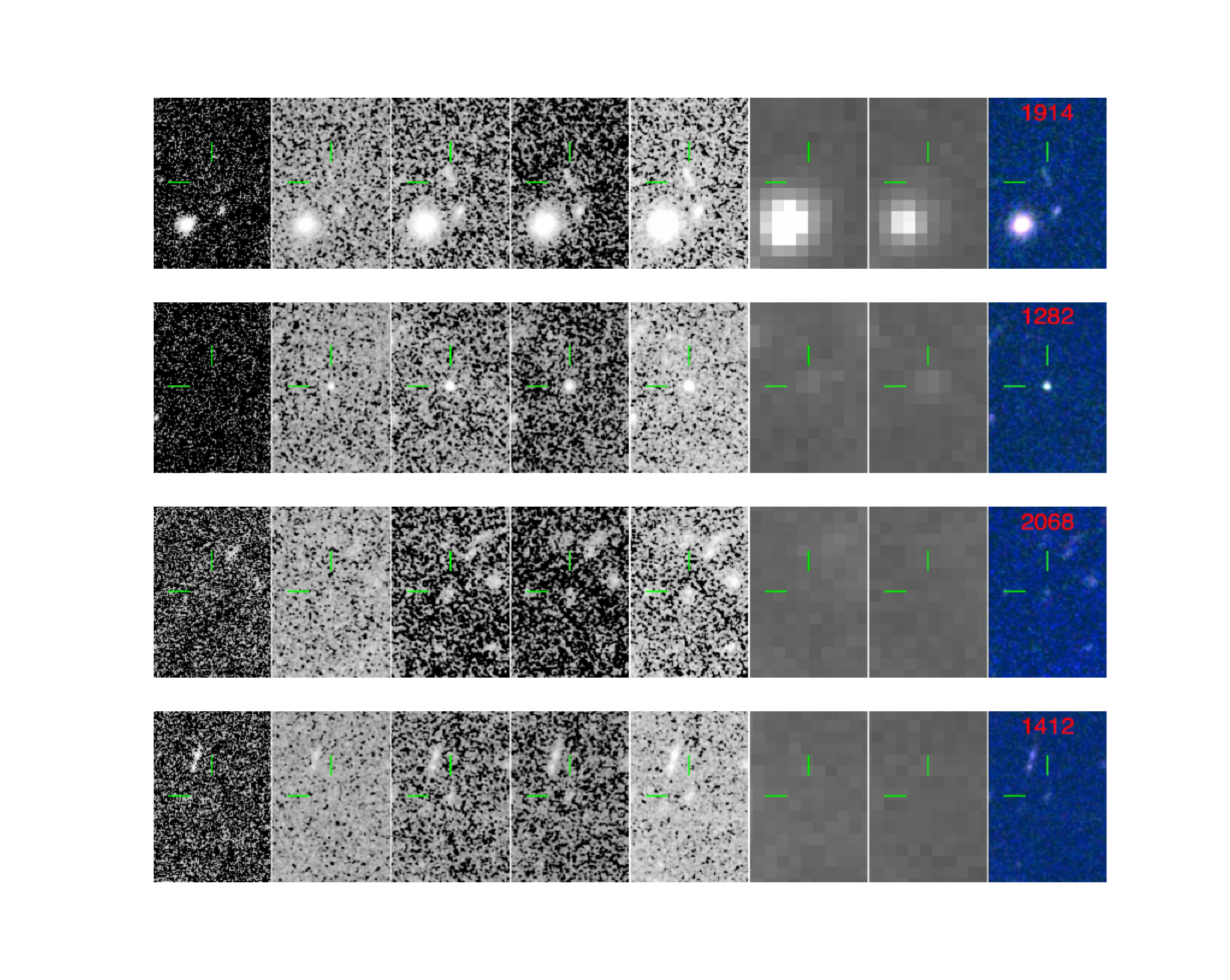}
\caption{Four primary F814W-dropout candidates behind RCS2327. From left to right: Combined $\mathrm{F435W} + \mathrm{F814W}$ optical image, F098M, F125W, F160W, combined $\mathrm{F098M} +\mathrm{F125W}+ \mathrm{F160W}$ infrared image, IRAC [3.6], IRAC [4.5] , and rgb image where b = F098M, g = F125W, and r = F160W.  Dropout IDs are shown in red on the rgb cutouts. All cutouts are $9''\times5.5''$. }    
\label{fig:drops}      
\end{figure*}


The F814W-dropout search area is limited by the size of the WFC3/IR FOV, which is $136''\times123''$, compared to the $240''\times240''$ grid used in the lens model. To conduct the search for F814W-dropouts, we closely follow the method described by \citet{coe06}, \citet{stark09}, and \citet{hall12}. Specifically, we run SExtractor in dual image mode, using a combined $\mathrm{F098M} + \mathrm{F125W} + \mathrm{F160W}$ infrared image as the detection image. For each object detected in the combined infrared image, we measure ISO and AUTO magnitudes in each of the five HST filter images (Table~\ref{table:hstfilts}). Huang et al. (2015, in preparation) show that ISO colors perform best over the magnitude range in which we are interested ($25<m<28$). AUTO magnitudes, however, are more representative of the total magnitude. We therefore use ISO colors, but add to the ISO magnitude in each filter the constant term: MAG\_ISO\_F160W - MAG\_AUTO\_F160W to recover the total magnitude in each filter. Magnitude errors reported from SExtractor are scaled by a constant factor in each filter which is derived by Huang et al. (2015, in preparation). The method is similar to what is described by \citet{trenti11} to correct photometric errors in the presence of correlated noise.


\begin{figure*}
\centering
\includegraphics[clip, width=\linewidth,height=18cm]{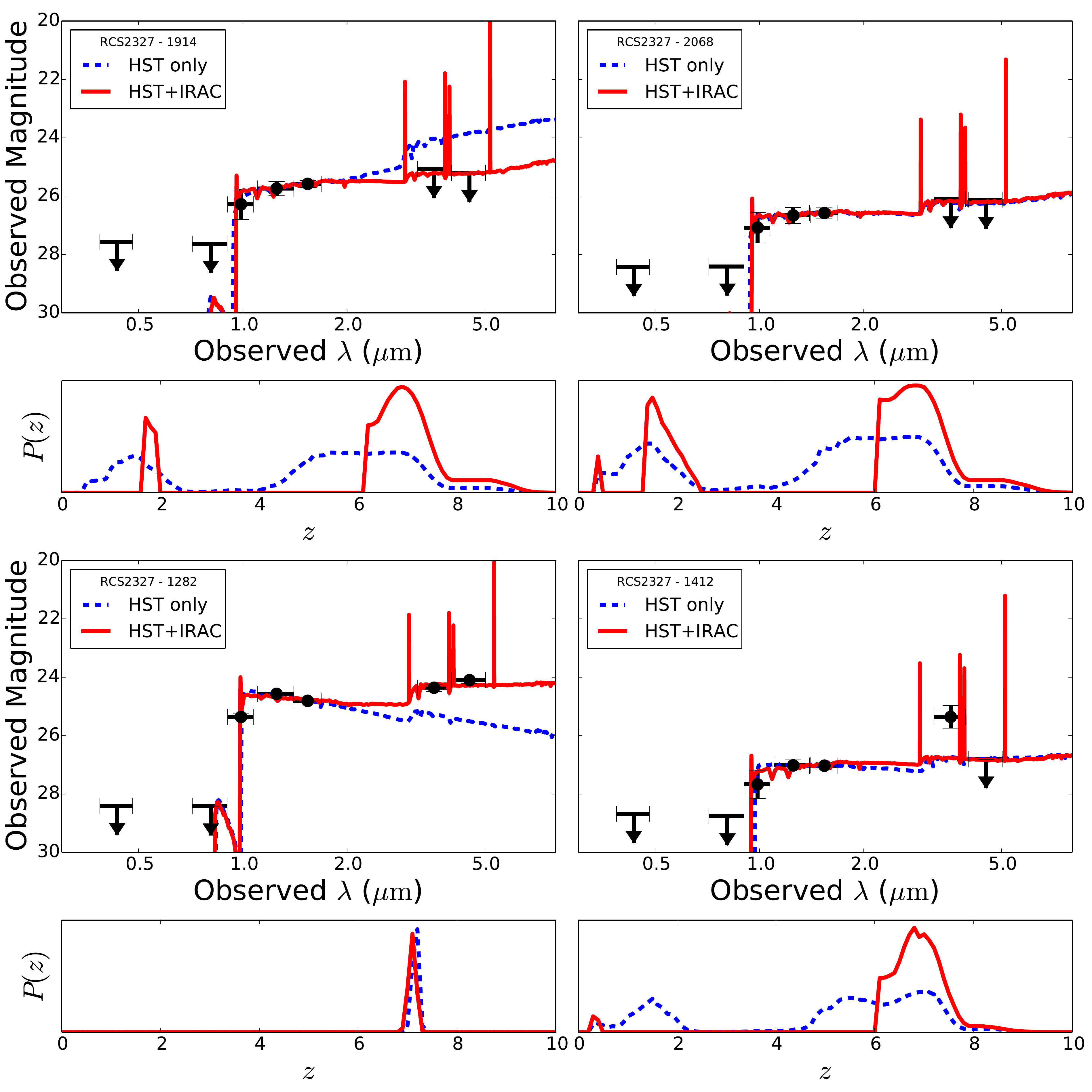}
\caption{Spectral energy distributions (top panels) and photometric redshift distributions (bottom panels) for the four primary F814W-dropout candidates found behind RCS2327. Shown are fits to the photometric data points using HST data only (dotted blue line) and using HST+ IRAC [3.6] and [4.5] data (solid red line). }    
\label{fig:dropSEDs}      
\end{figure*}


Here we describe the procedure for selecting F814W-dropouts, i.e., galaxies that show no significant flux in F814W or bluer bands. We employ two different color and $S/N$ selections and then compare the resulting two samples. Our first selection method is based on the color criteria used by \citet{bouwens11} and \citet{hall12}, where they targeted $z_{850}$ dropouts. Their cuts are also appropriate for efficiently eliminating contaminants from our selection, while simultaneously probing the region of color-color space occupied by $z\sim7$ starburst galaxies (Fig.~\ref{fig:dropcolors}). The first set of color criteria are:
\beq
\left\{
\begin{split}
\mathrm{F814W} - \mathrm{F125W} & > 1.0 + 0.4 (\mathrm{F125W} - \mathrm{F160W})  \\
\mathrm{F814W} - \mathrm{F125W} & > 1.0 \\
\mathrm{F125W} - \mathrm{F160W} & < 0.5 
\end{split}
\right.
\label{selection1}
\eeq

We visually inspect all candidates that fulfill the above color criteria. Objects that are clearly detected in F435W and F814W are discarded. We settle on 5 objects which show no significant flux above the background in both ACS filters.

Our second selection method is similar, but it uses automatic $S/N$ criteria in place of visual inspection. The color and $S/N$ criteria for this selection are:

\beq
\left\{
\begin{split}
& (\mbox{F814W} - \mbox{F125W}) >  1.1  \\
& (\mbox{F814W} - \mbox{F125W}) > (0.8 + 1.1 (\mbox{F125W} - \mbox{F160W}))  \\
& (\mbox{F814W} - \mbox{F125W}) > (0.4 + 1.1 (\mbox{F098M} - \mbox{F125W})) \\
& (\mbox{F098M} - \mbox{F125W}) < 1.25  \\
& (\mbox{F125W} - \mbox{F160W}) < 0.4  \\
& F814W > 25 \\
& S/N \leq 2.0 \, \mbox{in F435W} \\
& S/N \leq 2.0 \, \mbox{in F814W} \\
& S/N \leq 5.0 \, \mbox{in F125W} \\
& S/N \leq 5.0 \, \mbox{in F160W }
\end{split}
\right.
\eeq

7 objects fulfill these criteria, 4 of which were independently identified by the first selection method. We will refer to the four objects that were selected using both methods as the primary dropout candidates. To be more inclusive in our spectroscopic sample, we include the four additional candidates that were selected by one method but not by both. We will refer to these four as the secondary dropout candidates.  We show cutouts of the four primary dropout candidates in Fig.~\ref{fig:drops}. After the spectroscopic data were taken, the photometric errors were rescaled as described by Huang et al. (2015, in preparation). This does not significantly change the photometry of the dropouts. However, we note that in Fig.~\ref{fig:dropcolors}, we show the updated photometry for the dropout candidates, while showing the original color criteria from the first selection method to select the spectroscopic targets. 

Photometric redshifts were not used as part of the dropout selection criteria. They were obtained after the selection using Le Phare (Table~\ref{table:lbgs}). The full photometric redshift distributions are shown in Fig.~\ref{fig:dropSEDs}. All four dropouts favor $z\sim7$ solutions, regardless of whether we include IRAC [3.6] and [4.5] data. We examine the photometric redshift distributions more closely in Section~\ref{sec:limits}. 

Fig.~\ref{fig:rgbarcs} shows the location of the critical curve for a source at $z=7$. $3/4$ of our dropout candidates are located outside of the critical curve. RCS2327 - 1412 is located inside the critical curve at its photometric redshift, $z=6.8$. We check for an image on the opposite side of the critical curve at the expected location predicted by the model, but we do not detect it in the HST data. Because we lack strong lensing constraints near the location of RCS2327 - 1412, our model is more uncertain in this area. Therefore, the model may not be able to accurately predict counter-images at this position. An additional counter-image is predicted on the other side of the cluster, but would be too faint to detect in our current HST data.

As a result of their large magnifications, all four of the primary dropout candidates are intrinsically fainter than $L_{\star}$ at $z=6.8$\footnote{We use the characteristic luminosity, $L_{\star}$,  determined by \citealp{bouwens11}. }. We measure their ages and intrinsic stellar masses. Errors in these quantities are obtained by resampling the photometry from a gaussian of location and scale equal to the measured magnitude and magnitude error, respectively. We do this 1000 times and rerun Le Phare on each resample.


\begin{deluxetable*}{lccccccccc}
\tablecolumns{10}
\tablecaption{\bf Dropout Candidates \label{table:lbgs} } 
\tablehead{\colhead{ID} & \colhead{$\alpha_{J2000}$} & \colhead{$\delta_{J2000}$} & \colhead{F160W} & \colhead{ $\mu$} & \colhead{$z_\mathrm{phot}$} & \colhead{$L / L_{\star}$ } & \colhead{$log_{10} (Age/yr) $} & \colhead{ $log_{10}(M/M_{\odot}) $} & $\sigma_{W}$/{\AA} }
\medskip
\startdata
\vspace{0.2cm}
RCS2327-1914 & 351.878006 & -2.0875638 & 25.6 & $4.55^{+0.45}_{-0.24}$  & 6.9 & $0.22^{+0.03}_{-0.02}$ & $8.22^{+0.02}_{-0.03}$ & $9.08^{+0.38}_{-0.33}$ & 2.2 \\ 
\vspace{0.2cm}
RCS2327-1282 & 351.880689 & -2.0763742 & 24.8 & $4.11^{+0.46}_{-0.37}$  & 7.1 & $0.49^{+0.05}_{-0.05} $ & $8.77^{+0.02}_{-0.04}$ & $9.39^{+0.04}_{-0.07}$ &  1.0 \\ 
\vspace{0.2cm}
RCS2327-2068 & 351.856179& -2.0933244 & 26.6 & $10.42^{+2.01}_{-1.52}$  & 6.8 & $0.04^{+0.01}_{-0.01} $ & $8.25^{+0.01}_{-0.24}$ & $8.28^{+0.32}_{-0.25}$  &  4.9 \\ 
\vspace{-0.2cm}
RCS2327-1412 & 351.856215& -2.0785714 & 27.0 & $10.07^{+0.72}_{-0.43}$ & 6.8 & $0.027^{+0.003}_{-0.002}$ & $8.05^{+0.06}_{-0.16}$ & $8.58^{+0.00}_{-0.71}$ & 7.7 \\ 
\enddata
\tablecomments{Properties of the four primary F814W-dropout candidates. The column F160W lists is the apparent magnitude in that filter. Photometric redshifts, $z_{\mathrm{phot}}$ were obtained from the Le Phare $Z_{ML}$ keyword. The full photometric redshift distributions are shown in Fig.~\ref{fig:dropSEDs}. The characteristic luminosity, $L_{\star}$, was determined from the luminosity function at $z=6.8$ determined by \citealp{bouwens11}. $log_{10}(M/M_{\odot})$ represents the ``demagnified'' stellar mass.  The final column represents the median rest frame {\lya} equivalent width noise ($1\sigma$). All errors reflect $68\%$ confidence.} 
\end{deluxetable*}


\section{MOSFIRE NIR Spectroscopic Follow-up of F814W-Dropout Candidates}{\label{sec:limits}}

We observed all 8 F814W-dropout candidates (primary and secondary) with MOSFIRE Y and H-band for 3 hours. We performed a visual search for emission lines in all reduced 2-dimensional spectra. We subject each potential emission line we visually identify to a series of tests. 

\begin{figure}
        \centering
        \includegraphics[width=\linewidth]{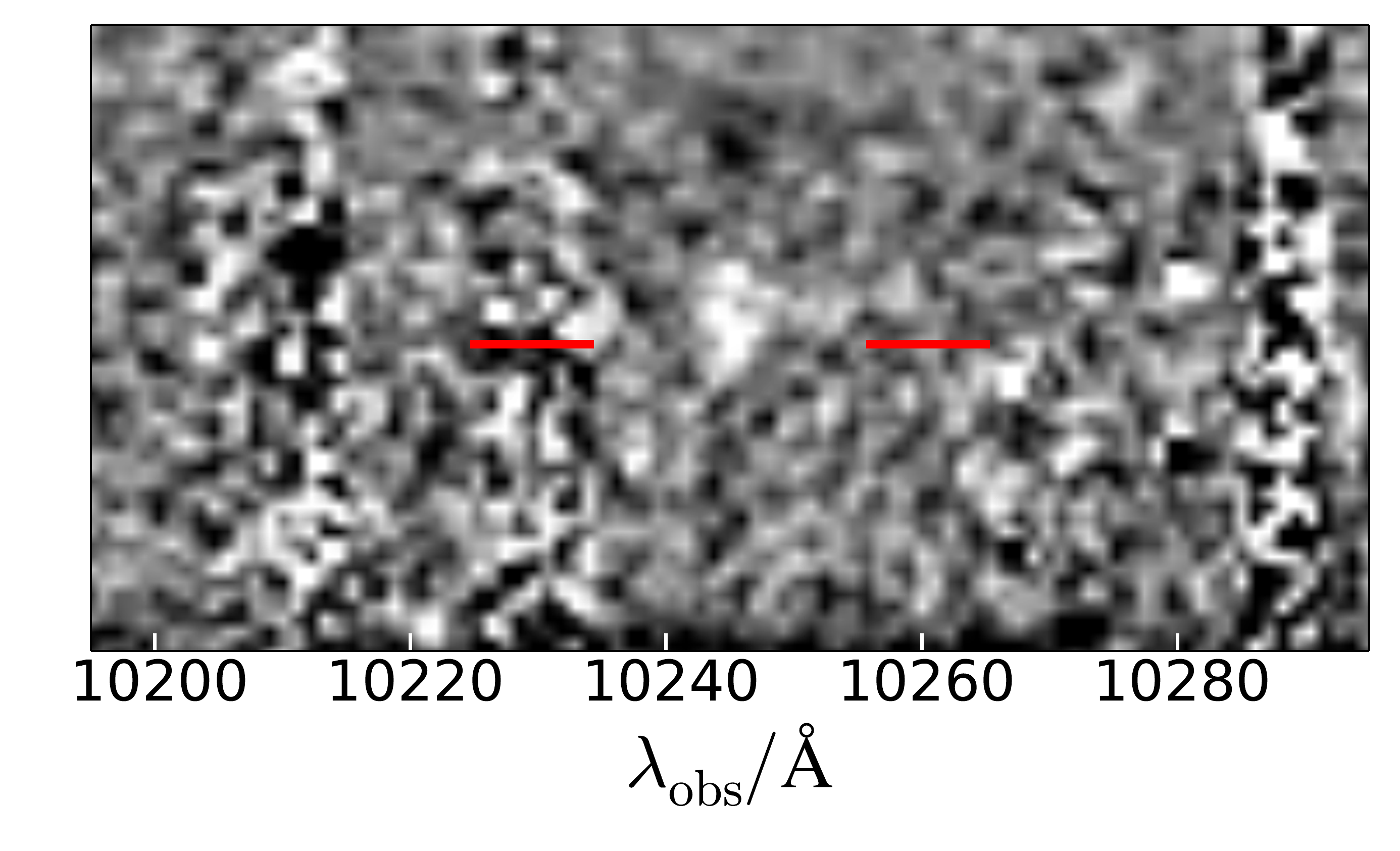}      
        \caption{Line-like feature at $\lambda_{\mathrm{obs}} = 10245$ {\AA} in 3 hr combined MOSFIRE Y-band spectrum of dropout candidate RCS2327 - 843. If the feature is {\lya}, which we find unlikely, the galaxy is at $z=7.42$. The red horizontal lines indicate the expected vertical position in the slit from the broadband filter. The feature is well separated from nearby sky line residuals. The portion of the spectrum shown is 9$''$ in the vertical direction.}
        \label{fig:lya_emission}
\end{figure}

\begin{enumerate}
	\item The line is not significantly offset with respect to the continuum position.
	\item The line possesses two negative residuals at the positions expected from the nod pattern, each with $\simeq$ half the strength of the line itself. 
	\item The line FWHM is larger than or equal to the seeing FWHM in the spatial direction and is larger than or equal to the instrumental FWHM along the spectral axis. 
	\item The line flux is comparable in data from individual nights. 
	\item The line is not due to a known sky emission line. 	
	\item The redshift suggested by the wavelength of the line falls within the $95\%$ confidence limits of the photometric redshift distribution calculated using Le Phare. 
\end{enumerate}
After visually inspecting the 2D spectra of all primary dropout candidates, we do not detect any viable emission lines. We do, however, detect a line feature at $3.6\sigma$ in the combined Y-band data of object RCS2327 - 843, a secondary dropout candidate (Fig.~\ref{fig:lya_emission}). The feature is centered at 10245 {\AA}, which, if {\lya}, would mean the galaxy is at $z=7.42$. Despite passing tests 1, 3 and 5, the line fails tests 2, 4 and 6. Upon inspecting the reduced spectra of this object from individual nights, we only see a significant detection in the last night of observations, Dec 18. Furthermore, the photometric redshift distribution rules out $z=7.42$ at greater than $98\%$ confidence, even if we subtract off the line flux observed from the flux measured in F098M in HST before computing the photometric redshift. The photometric redshift distribution peaks at $z=6.7$ and similarly prohibits [OII]$\la\la$3726-3729 at $z=1.74$, the case in which the photometric break is the Balmer break instead of the Lyman break. The night in which the feature appears had the worst seeing (Table~\ref{table:mosobs}). We therefore  conclude that it is very unlikely that the feature is an emission line from RCS2327 - 843. RCS2327 - 843 is well isolated in HST data; the closest object that we find is at a distance greater than $3''$ from the position of the emission line, whereas the seeing was only $\simeq1.20''$ during observations that night. The feature is therefore unlikely to be an emission line produced by a neighboring object.

In addition to visually searching through all candidate dropout spectra, we conduct an automatic search by extracting 1D signal-to-noise spectra from the 2-dimensional reduced spectra. With a $\mbox{$S/N$}=4$ detection threshold in each separate dropout spectrum, 10 features are detected in the spectra of the four primary dropouts. None of these features pass tests 1-6, however.

We calculate {\lya} flux limits in Y-band following the flux calibration and 1-dimensional extraction procedures described in Section~\ref{sec:data}. Not accounting for magnification, we measure median $5\sigma$ {\lya} flux limits of $0.5-0.6 \times 10^{-17}\, \mathrm{erg \,\, s^{-1} cm^{-2} }$, in agreement with the limits observed by \citet{treu13}, who took observations with similar seeing and exposure times. After accounting for magnification, we achieve limits of $5-14 \times 10^{-19}\, \mathrm{erg \,\, s^{-1} cm^{-2}} $. Rest frame EW limits are obtained from the flux limits relative to the HST continuum measured in F098M (Table~\ref{table:lbgs}). Since EW is a relative flux measurement, it is unaffected by lensing magnification. We measure median rest frame EW limits of $1-8$ ($1\sigma$) for the primary four dropout candidates.

The fact that we do not detect significant {\lya} emission is unsurprising given our small sample size and the small ($\lesssim10\%$) {\lya} fraction at $z\simeq7$ and above (\citealp{schenker12}, \citealp{treu12}, \citealp{bradac12}, \citealp{finkelstein13} \citealp{treu13}, \citealp{pentericci14}, \citealp{schenker14}). This is in contrast to a {\lya} fraction of $\simeq50\%$ at $z\simeq6$ (\citealp{stark10}). The sharp decline in the {\lya} fraction from $z=6$ to $z=7$ is typically attributed to the attenuation of {\lya} due to the rising abundance of neutral hydrogen in the IGM (\citealp{treu13} and \citealp{schenker14}). \citet{mesinger15} show with simulations of the IGM during the reionization process that such a sharp decline is unlikely due to IGM attenuation alone. Alternative explanations include evolution in galaxy properties (\citealp{dijkstra14}) or a rising abundance of absorption systems (\citealp{bolton13}). 

\citet{stark14a} suggest using rest-frame UV lines, such as CIII]$\la$1909{\AA} for redshift confirmation in place of {\lya}. While typically a weaker line than {\lya}, CIII]$\la$1909{\AA} is not resonantly scattered by neutral hydrogen. CIII] is of particular interest in our sample because we probe the sub-$L_{\star}$ population, from which 90\% of the total UV luminosity density comes, at least at $z\simeq2$ (\citealp{reddy09}, \citealp{oesch10b}, \citealp{alavi14}, \citealp{stark14a}). We conduct similar visual and automatic searches for the CIII]$\la$1909{\AA} doublet in our H-band spectra of the dropout candidates. We run tests (1-6) on potential emission lines in H-band as we did in Y-band. Again, we see no convincing emission lines. 

\subsection{Contamination}

M, L, and T dwarfs have similar colors to $z\sim6-7$ starburst galaxies (Fig.~\ref{fig:dropcolors}). While RCS2327 is well removed from the galactic plane ($b\simeq-58$), we nonetheless investigate M, L, and T dwarfs as a source of contamination in detail. We inspect each of the dropout candidates in HST to determine whether it is resolved. All of the candidates are clearly resolved except RCS2327-1282, the brightest object in our sample. We attempt to fit its photometry to M, L and T dwarf spectra using the SpeX prism libraries\footnote{\url{http://pono.ucsd.edu/~adam/browndwarfs/spexprism/}}. We are unable to find a good fit to any of the dwarf SEDs. The SpeX libraries only extend out to the near-IR, but RCS2327-1282 is detected at $>10 \sigma$ in both IRAC [3.6] and [4.5]. At these wavelengths, we rely on a sample of M, L and T dwarfs studied by \citet{patten06} for the comparison. We compare the $\mathrm{F125W} - [3.6]$ and $[3.6]-[4.5]$ colors of RCS2327-1282 with the colors of the dwarfs, finding consistency with their T dwarfs only. Therefore, we cannot rule out the possibility that RCS2327-1282 is a T dwarf contaminant. 

Another source of contamination in our sample is the population of elliptical galaxies at $z\sim1$. The addition of IRAC [3.6] and [4.5] to HST data can help lift the degeneracy between $z<2$ galaxies and $z>6$ starburst galaxies. Fig.~\ref{fig:dropSEDs} illustrates this. In particular for candidates RCS2327 - 1914 and RCS2327 - 1412, for which we obtained IRAC limits only, high redshift solutions are much more favored once the IRAC data are included.

\section{Conclusion}
Our new gravitational lens model indicates that {\rcs} is an excellent cosmic telescope, as efficient as the Frontier Fields galaxy clusters. We discover 16 new multiple images consisting of 6 new systems. We identify a strong emission line at 15436 {\AA} in the MOSFIRE H-band spectrum of one arc, which is likely the [OIII]$\la$5007 emission line at $z=2.083$. The spectroscopic redshift suggested by the emission line is in concert with the tight constraints from the photometric redshift of the multiple image system.  

We find four highly magnified sub-$L_{\star}$ F814W-dropout candidates behind the cluster. Measuring their photometric redshifts, we find that they are likely at $z\sim7$. We also find four F814W-dropout candidates that fulfill less strict selection criteria. We search for {\lya} emission from all eight candidates using Keck/MOSFIRE. We identify a feature at $3.6\sigma$ at 10245 {\AA} in the Y-band spectrum of one the secondary dropout candidates, but conclude that it is unlikely a true emission line. We see no convincing emission lines in the four primary dropout candidates, reaching down to median $5\sigma$ {\lya} flux limits of $5-14 \times 10^{-19}\, \mathrm{erg \,\, s^{-1} cm^{-2}}$. We also measure median $1\sigma$ {\lya} rest frame EW limits of $1-8$. Because {\lya} emission is affected differently by the presence of neutral hydrogen in the IGM in different reionization scenarios, the {\lya} EW distribution at $z\gtrsim7$ can be used to distinguish between models of reionization. The EW limits we measured from our MOSFIRE follow-up will be used as part of a larger sample obtained from dropout searches behind all SURFSUP clusters to help better understand the nature of reionization.    

\section{Acknowledgements}
Observations were carried out using Spitzer Space Telescope, which is operated by the Jet Propulsion Laboratory, California Institute of Technology under a contract with NASA. Also based on observations made with the NASA/ESA Hubble Space Telescope, obtained at the Space Telescope Science Institute, which is operated by the Association of Universities for
Research in Astronomy, Inc., under NASA contract NAS5-26555 and NNX08AD79G and ESO-VLT telescopes. Support for this work was provided by NASA through
a Spitzer award issued by JPL/Caltech. This work was supported by NASA Headquarters under the NASA Earth and Space Science Fellowship Program - Grant ASTRO14F-0007. We also acknowledge support from HST-AR-13235, HST-GO-13177, and special funding as part of the HST Frontier Fields program conducted by STScI. TS acknowledges support from the German Federal Ministry of Economics and Technology (BMWi) provided through DLR under project 50 OR 1308. TT acknowledges support by the Packard Fellowship. HH is supported by the Marie Curie IOF 252760, by a CITA National Fellowship, and the DFG Emmy Noether grant Hi 1495/2-1. SA and AvdL acknowledge support by the U.S. Department of Energy under contract number DE-AC02-76SF00515 and by the Dark Cosmology Centre (DARK) which is funded by the Danish National Research Foundation.

This research has benefitted from the SpeX Prism Spectral Libraries, maintained by Adam Burgasser at \newline \url{http://pono.ucsd.edu/~adam/browndwarfs/spexprism/}. We would like to thank Chris Fassnacht for sharing his spectroscopic analysis knowledge and tools. Some of the data presented herein were obtained at the W.M. Keck Observatory, which is operated as a scientific partnership among the California Institute of Technology, the University of California and the National Aeronautics and Space Administration. The Observatory was made possible by the generous financial support of the W.M. Keck Foundation. The authors wish to recognize and acknowledge the very significant cultural role and reverence that the summit of Mauna Kea has always had within the indigenous Hawaiian community.  We are most fortunate to have the opportunity to conduct observations from this mountain.

\bibliography{bibliogr_highz}
\bibliographystyle{apj}

\end{document}